\input harvmac
\input epsf

\newcount\figno
\figno=0
\def\fig#1#2#3{
\par\begingroup\parindent=0pt\leftskip=1cm\rightskip=1cm\parindent=0pt
\baselineskip=11pt
\global\advance\figno by 1
\midinsert
\epsfxsize=#3
\centerline{\epsfbox{#2}}
\vskip 12pt
{\bf Fig. \the\figno:} #1\par
\endinsert\endgroup\par
}
\def\figlabel#1{\xdef#1{\the\figno}}
\def\encadremath#1{\vbox{\hrule\hbox{\vrule\kern8pt\vbox{\kern8pt
\hbox{$\displaystyle #1$}\kern8pt}
\kern8pt\vrule}\hrule}}

\overfullrule=0pt

\noblackbox
\parskip=1.5mm

\def\Title#1#2{\rightline{#1}\ifx\answ\bigans\nopagenumbers\pageno0
\else\pageno1\vskip.5in\fi \centerline{\titlefont #2}\vskip .3in}

\font\caps=cmcsc10

\noblackbox
\parskip=1.5mm

  
\def\npb#1#2#3{{\it Nucl. Phys.} {\bf B#1} (#2) #3 }
\def\plb#1#2#3{{\it Phys. Lett.} {\bf B#1} (#2) #3 }
\def\prd#1#2#3{{\it Phys. Rev. } {\bf D#1} (#2) #3 }
\def\prl#1#2#3{{\it Phys. Rev. Lett.} {\bf #1} (#2) #3 }

\def\ijmpa#1#2#3{{\it Int. J. Mod. Phys.} {\bf A#1} (#2) #3 }
\def\jmp#1#2#3{{\it J. Math. Phys.} {\bf #1} (#2) #3 }
\def\cmp#1#2#3{{\it Commun. Math. Phys.} {\bf #1} (#2) #3 }

\def\bb#1{{\tt hep-th/#1}}

\def\heph#1{{\tt hep-ph/#1}}
\def\mathph#1{{\tt math-ph/#1}}

\def\npps#1#2#3{{\it Nucl. Phys. Proc. Suppl. } {\bf #1} (#2) #3 }

\def\jhep#1#2#3{{\it JHEP} {\bf #1} (#2) #3 }
\def\cm#1{{\tt cond-mat/#1}}


\def\CA{{\cal A}} \def\CC{{\cal C}} \def\CF{{\cal F}} \def\CG{{\cal G}}

\def\CN{{\cal N}}  \def\CW{{\cal W}} \def\CV{{\cal V}}

\def\sumint{\hbox{$\sum$}\!\!\!\!\!\!\int}


\def\dj{\hbox{d\kern-0.347em \vrule width 0.3em height 1.252ex depth
-1.21ex \kern 0.051em}}

\def\half{{1\over 2}\,}

\def\ket{\rangle}

\def\Dirac{\,\raise.15ex\hbox{/}\mkern-13.5mu D}
\def\dirac{\,\raise.15ex\hbox{/}\kern-.57em \partial}
\def\shalf{{\ifinner {\scriptstyle {1 \over 2}}\else {1 \over 2} \fi}} 

\lref\rpert{T. Filk, \plb{376}{1996}{53\semi}
J.C. V\'arilly and J.M. Gracia-Bond\'{\i}a, \ijmpa{14}{1999}{1305}
(\bb{9804001})\semi
C.P. Mart\'{\i}n and D. S\'anchez-Ruiz, \prl{83}{1999}{476} (\bb{9903077})\semi
M.M. Seikh-Jabbari, \jhep{06}{1999}{015} (\bb{9903107})\semi
T. Krajewski and R. Wulkenhaar, {\it Perturbative quantum gauge fields
on the noncommutative torus}, \bb{9903187}\semi
D. Bigatti and L. Susskind, {\it Magnetic fields, branes and noncommutative
geometry}, \bb{9908056}\semi
I.Ya. Aref'eva, D.M. Belov and A.S. Koshelev, \plb{476}{2000}{431} 
(\bb{9912075})\semi
M. Hayakawa, {\it Perturbative analysis on infrared aspects of noncommutative QED on
${\bf R}^{4}$}, \bb{9912094}; {\it Perturbative analysis on infrared and ultraviolet 
aspects of noncommutative QED on ${\bf R}^{4}$}, \bb{9912167}\semi
I.Ya. Aref'eva, D.M. Belov and A.S. Koshelev, {\it A note
on UV/IR for noncommutative complex scalar field}, \bb{0001215}\semi
A. Matusis, L. Susskind and N. Toumbas, {\it The IR/UV connection in non-commutative
gauge theories}, \bb{0002075}.}
\lref\marti{T. Banks, M. R. Douglas, G. T. Horowitz and E. Martinec,
hep-th/9808016. } 
\lref\rzac{D.B. Fairlie, P. Fletcher and C.K. Zachos, \jmp{31}{1990}{1088.}}
\lref\mac{A.J. Macfarlane, A. Sudbery and P.H. Weisz, \cmp{11}{1968}{77.}}
\lref\kphd{T. Krajewski, {\it G\'eom\'etrie non commutative et interactions
fondamaentales}, Ph.D. Thesis. (\mathph{9903047})}
\lref\rj{M.M. Seikh-Jabbari, \jhep{06}{1999}{015.} (\bb{9903107})}
\lref\rkw{T. Krajewski and R. Wulkenhaar, {\it Perturbative quantum gauge fields
on the noncommutative torus}, Preprint CPT-99/P.3794. (\bb{9903187})}
\lref\rthooft{G. 't Hooft, \npb{153}{1979}{141;} \cmp{81}{1981}{267.}}
\lref\kap{J.I. Kapusta, \npb{148}{1979}{461.}}
\lref\bs{D. Bigatti and L. Susskind, {\it Magnetic fields, branes and 
nocommutative geometry}, Preprint SU-ITP-99-39. (\bb{9908056})}
\lref\rncym{T. Filk, \plb{376}{1996}{53.}}
\lref\rmagf{P. van Baal, \cmp{85}{1982}{529\semi}
J. Troost, {\it Contant field strenghts on $T^{2n}$}, Preprint VUB-TENA-99-04.
(\bb{9909187})}
\lref\br{J.L.F. Barb\'on and E. Rabinovici, \jhep{04}{1999}{015.} (\bb{9910019})}
\lref\rsw{N. Seiberg and E. Witten, \jhep{09}{1999}{032.} (\bb{9908142})}
\lref\niels{T. Harmark and N. Obers, {\it Phase structure of field theories
and spinning brane bound states}, Preprint NBI-HE-99-47. (\bb{9911169})}
\lref\rmr{J.M. Maldacena and J.G. Russo, \jhep{09}{1999}{25}. (\bb{9908134})}
\lref\rhio{A. Hashimoto and N. Itzhaki, \plb{465}{1999}{142.} (\bb{9907166})}
\lref\rcds{A. Connes, M. Douglas and A. Schwarz, \jhep{02}{1998}{003}. (\bb{9711162})}
\lref\wsft{E. Witten, \npb{268}{1986}{253.}}
\lref\rps{B. Pioline and A. Schwarz, \jhep{08}{1999}{021.} (\bb{9908019})}
\lref\rdh{M.R. Douglas and C.M. Hull, \jhep{02}{1998}{008.} (\bb{9711165})}
\lref\qhe{J. Bellisard, A. van Elst and H. Schulz-Baldes, {\it The non-commutative
geometry of the quantum Hall effect}, \cm{9411052}.}
\lref\ac{A. Connes and J. Lott, \npps{18}{1990}{29.}}
\lref\vm{M.A. V\'azquez-Mozo, \prd{60}{1999}{106010.} (\bb{9905030})}
\lref\ft{A. Fotopoulos and T.R. Taylor, \prd{59}{1999}{061701.} (\bb{9811224})}
\lref\sjr{C. Kim and S.-J. Rey, {\it Thermodynamics of large-N super Yang-Mills and the
AdS/CFT correspondence}, Preprint SNUST-99-005 (\bb{9905205}).}
\lref\an{A. Nieto and M.H.G. Tytgat, {\it Effective field theory approach to N=4 
supersymmetric Yang-Mills at finite temperature}, Preprint CERN-TH-99-153 (\bb{9906147}).}
\lref\raz{P. Arnold and C. Zhai, \prd{50}{1994}{7603} (\heph{9408276});
\prd{51}{1995}{1906} (\heph{9410360})\semi
C. Zhai and B. Kastening, \prd{52}{1995}{7232} (\heph{9507380}).}
\lref\rcb{A. Connes, {\it Noncommutative Geometry}, Academic Press 1994.}
\lref\rcar{C.P. Mart\'{\i}n and D. S\'anchez-Ruiz, \prl{83}{1999}{476.} (\bb{9903077})}
\lref\rtoi{T. Toimela, \plb{124}{1983}{407.}}
\lref\rbn{E. Braaten and A. Nieto, \prd{53}{1996}{3421.} (\bb{9510408})}
\lref\ob{T. Harmark and N.A. Obers, {\it Phase structure of noncommutative
field theories and spinning brane bound states}, Preprint NBI-HE-99-47 (\bb{9911169}).}
\lref\arlh{A. Connes, {\it Non-commutative geometry and physics}, in: "Gravitation and 
Quantizations", Proceedings of the 1992 Les Houches Summer School. Eds. B. Julia and
J. Zinn-Justin. Elsevier 1995.}
\lref\rpp{J.C. V\'arilly and J.M. Gracia-Bond\'{\i}a, \ijmpa{14}{1999}{1305.} (\bb{9804001})} 
\lref\rhid{A. Hashimoto and N. Itzhaki, {\it On the hierarchy between non-commutative and
ordinary supersymmetric Yang-Mills}, Preprint NSF-ITP-99-133, (\bb{9911057}).}
\lref\aoj{M. Alishahiha, Y. Oz and M.M. Seikh-Jabbari, {\it 
Supergravity and large N noncommutative field theories}, \jhep{11}{1999}{007.}
(\bb{9909215})}
\lref\rhaya{M. Hayakawa, {\it Perturbative analysis on infrared aspects of noncommutative
QED on ${\bf R}^{4}$}, (\bb{9912094}).}
\lref\rp{S. Minwalla, M. Van Raamsdonk and N. Seiberg, {\it Noncommutative perturbative
dynamics}, (\bb{9912072}).}
\lref\aref{I.Ya. Aref'eva, D.M. Belov and A.S. Koshelev, \plb{476}{2000}{431} 
(\bb{9912075}).}
\lref\rgc{H. Garc\'{\i}a-Compean, \npb{541}{1999}{651.} (\bb{9804188})}
\lref\rtexas{W. Fischler, J. Gomis, E. Gorbatov, A. Kashani-Poor, S. Paban and
P. Pouliot, {\it Evidence for winding states in noncommutative quantum field
theory}, \bb{0002067}.}
\lref\rrase{M. Van Raamsdonk and N. Seiberg, \jhep{03}{2000}{035.} (\bb{0002186})}
\lref\rgms{R. Gopakumar, S. Minwalla and A. Strominger, {\it Noncommutative
solitons}, \bb{0003160}.}
\lref\rbil{A. Bilal, C.-S. Chu and R. Russo, {\it String theory and noncommutative
field theories at one loop}, \bb{0003180}.}
\lref\rjaume{J. Gomis, M. Kleban, T. Mehen, M. Rangamani and S. Shenker, 
{\it Noncommutative gauge dynamics from the string worldsheet}, \bb{0003215}.}
\lref\rkl{Y. Kiem and S. Lee, {\it 
UV/IR mixing in noncommutative field theory via open string loops}, 
\bb{0003145}}
\lref\rad{O. Andreev and H. Dorn, {\it Diagrams of Noncommutative $\Phi^3$ Theory
from String Theory}, \bb{0003113}.}
\lref\raroz{A. Rajaraman and M. Rozali, {\it Noncommutative gauge theory,
divergences and closed strings}, \bb{0003227}.}
\lref\rmvm{M.A. V\'azquez-Mozo, \plb{388}{1996}{494,} (\bb{9607052}).}
\lref\rtext{W. Fischler, E. Gorbatov, A. Kashani-Poor, R. McNees, S. Paban
and P. Pouliot, {\it The interplay between $\theta$ and $T$}, \bb{0003216}.}
\lref\rcaioh{R.-G. Cai and N. Ohta, \jhep{003}{2000}{009.} (\bb{0001213})}
\lref\ravm{G. Arcioni and M.A. V\'azquez-Mozo, \jhep{01}{2000}{028.}
(\bb{9912140})}
\lref\rkbo{J.I. Kapusta, {\it Thermal Field Theory}, Cambridge 1989.}
\lref\rcaiot{R.-G. Cai and N. Ohta, {\it On the thermodynamics of large-N
noncommutative super Yang-Mills theory}, \bb{9910092}.}
\lref\rlmi{H. Liu and J. Michelson, {\it Stretched strings in noncommutative
field theory}, \bb{0004013}.}
\lref\rareft{I.Ya. Aref'eva, D.M. Belov and A.S. Koshelev, {\it A note
on UV/IR for noncommutative complex scalar field}, \bb{0001215}.}
\lref\rnonp{A. Hashimoto and N. Itzaki, \plb{465}{1999}{142} 
(\bb{9907166})\semi
J.M. Maldacena and J.G. Russo, \jhep{09}{1999}{025} (\bb{9908134})\semi
M. Alishahiha, Y. Oz and M.M. Seikh-Jabbari, \jhep{11}{1999}{007}
(\bb{9909215})\semi
A. Hashimoto and N. Itzaki, \jhep{12}{1999}{007} (\bb{9911057})\semi
T. Harmark and N.A. Obers, \jhep{03}{2000}{024.} (\bb{9911169})}
\lref\raw{J.J. Atick and E. Witten, \npb{310}{1988}{291.}}
\lref\rvmdb{M.A. V\'azquez-Mozo, \plb{388}{1996}{494.} (\bb{9607052})}
\lref\rgll{Joaquim Gomis, K. Landsteiner and E. L\'opez, work in progress.}
\lref\rcallan{A. Abouelsaood, C.G. Callan, C.R. Nappi and S.A. Yost, 
\npb{280}{1987}{599.}}


\baselineskip=14pt

\line{\hfill CERN-TH/2000-114}
\line{\hfill DFTT 15/2000}
\line{\hfill SPIN-2000/13}
\line{\hfill UB-ECM-PF-00/06}
\line{\hfill ITFA-2000-08}
\line{\hfill {\tt hep-th/0004080}}
\vskip 0.5cm

\Title{\vbox{\baselineskip 12pt\hbox{}
 }}
{\vbox {\centerline{On the Stringy Nature of Winding Modes in}
\vskip10pt
\centerline{Noncommutative Thermal Field Theories}
}}

\centerline{$\quad$ {\caps G. Arcioni~$^{a,b}$,
J.L.F. Barb\'on~$^{c,}$\foot{ On leave
from: Departamento de F\'{\i}sica de Part\'{\i}culas,
Universidad de Santiago de Compostela, Spain.},
Joaquim Gomis~$^{d,c}$ and M.A. V\'azquez-Mozo~$^{e,b}$ 
}}
\vskip0.2cm

\centerline{{\sl $^a$ Dipartimento di Fisica Teorica, Universit\`a di Torino,
via P. Giuria 1}}
\centerline{{\sl  I-10125 Torino, Italy {\rm and} INFN Sezione di Torino}}
\centerline{{\tt arcioni@to.infn.it}}
\vskip0.2cm

\centerline{{\sl $^b$ Spinoza Instituut, 
Universiteit Utrecht, Leuvenlaan 4, 3584 CE Utrecht, The Netherlands}}  
\centerline{{\tt g.arcioni@phys.uu.nl, M.Vazquez-Mozo@phys.uu.nl}}

\vskip0.2cm 

\centerline{{\sl $^c$ Theory Division, CERN, 
 CH-1211 Geneva 23, Switzerland}}
\centerline{{\tt barbon@mail.cern.ch, gomis@mail.cern.ch}} 

\vskip0.2cm

\centerline{{\sl $^d$ Departament ECM, Facultat de Fisica, Universitat de Barcelona}}
\centerline{{\sl {\rm and} Institut de Fisica d'Altes Energies, Diagonal 647, E-08028 
Barcelona, Spain}}
\centerline{{\tt gomis@ecm.ub.es}}

\vskip0.2cm

\centerline{{\sl $^e$ Instituut voor Theoretische Fysica, Universiteit van
Amsterdam}}
\centerline{{\sl Valckenierstraat 65, 1018 XE Amsterdam, The Netherlands}} 
\centerline{{\tt vazquez@wins.uva.nl}}

\vskip0.2cm

\vskip 2.5cm

               
\noindent CERN-TH/2000-114 

\vfill

\centerline{\bf ABSTRACT}

 \vskip 0.3cm

\noindent 
We show that thermal noncommutative field theories admit a version
of `channel duality' reminiscent of open/closed string duality, where 
non-planar thermal loops can be replaced by an infinite tower of
tree-level exchanges of effective fields. These effective fields
resemble closed strings in three aspects: their mass spectrum is that
of closed-string winding modes, their interaction vertices contain
extra moduli, and they can be regarded as propagating in a higher-dimensional
`bulk' space-time.
 In noncommutative models that can be embedded in a D-brane,
we show the precise relation between the effective `winding fields' 
and closed strings propagating off the D-brane. The winding fields
represent the coherent coupling of the infinite tower of closed-string
oscillator states. We derive a sum rule that expresses this effective
coupling in terms of the elementary couplings of closed strings to the
D-brane. We furthermore clarify the relation between the effective
propagating
dimension of the winding fields and the true codimension of the D-brane.



\vskip 0.1cm


 
\Date{April 2000}



\baselineskip=14pt

\newsec{Introduction }

It has been realized  that noncommutative field theories (NCFT) emerge as  
effective field theories of
 string/M-theory compactifications in the presence of 
constant antisymmetric tensor fields \refs\rdh\refs\rcds\refs\rsw. 
This result has triggered a renewed interest in the study of 
both perturbative  \refs\rpert\refs\rp\refs\rrase\ and 
nonperturbative \refs\rnonp\ aspects of noncommutative field theories.
The stringy connection of NCFT opens up the interesting possibility of trying 
to understand some of their physical features by embedding them in string theory. 
In particular a number of NCFT have been obtained as a low-energy limit of 
open-string theories in $B$-field backgrounds 
\refs\rad\refs\rkl\refs\rbil\refs\rjaume\refs\raroz\refs\rlmi.
As a matter of example, one can try to understand 
the nonlocality inherent in these quantum field theories
in terms of string theory after an appropriate
low-energy limit is taken \refs\rsw.

Among the most intriguing features of NCFT is a peculiar
mixing between infrared and ultraviolet scales \refs\rp\refs\rrase. 
On physical grounds it can be understood as the result of the uncertainty
principle between two noncommuting spatial dimensions, since probing ultraviolet 
physics in one direction leads to infrared effects in the other. At a more 
technical level, this mixing reflects itself in the appearances of extra poles 
at zero momentum in some amplitudes in the limit where the ultraviolet cutoff 
is sent to infinity. The authors of Refs. \refs\rp\refs\rrase\ interpreted these poles 
as resulting from the interchange of
a new field $\psi$ with kinetic kernel $-\partial\circ\partial\equiv 
\partial_{\mu}(\theta^2)^{\mu\nu}\partial_{\nu}$.  
It would be very interesting to see if there is a stringy interpretation
for these particles.

One of the obvious ways of spotting stringy behavior
in NCFT would be to look at situations where the presence of extended
objects is made manifest, as for example studying these theories in spaces with
nontrivial topology or at finite temperature \refs\br\refs\rcaiot\refs\ravm. In Refs.
\refs\rtexas\refs\rtext\ it was pointed out that the two loop thermal partition function 
of some NCFT can be cast in a way that indicates the presence of states whose 
energy scales with the inverse temperature as $|\ell \beta|$, with $\ell$ some 
integer number. This would suggest that NCFT contains certain extended degrees 
of freedom that are able to wrap around the euclidean time.

In this paper we will try to understand whether some kind of winding modes
can be identified in noncommutative thermal perturbation theory, extending
on the work of \refs\rtexas. Actually, we shall see that the 
winding modes formally identified in thermodynamical 
quantities can be associated to effective fields with special propagators,
much in the same fashion as the $\psi$-fields of Refs. \refs\rp\refs\rrase.   
In fact, the UV/IR interpretation of these propagators is the same once
we realize that the temperature acts as an ultraviolet cutoff in the field
theory.

Therefore, this raises the question of whether the `winding fields' could
be interpreted as `off-brane' closed-string modes
 that survive the Seiberg--Witten
(SW) decoupling limit. We find that this expectation is not fulfilled, at least
in a literal sense. In particular, any closed-string picture amounts to 
the exchange of the infinite tower of closed-string excitations in the
bulk, 
and therefore it is not a very transparent way of describing the dynamics.
Instead, each winding field describes a sort of coherent exchange of an
infinite number of closed-string modes. One of our results is the
derivation of a sum rule for the effective coupling of the winding fields,
in terms of the elementary couplings of closed strings to a D-brane. 
In fact, the   interactions of these winding  fields  
 are {\it not} specified solely in terms of 
standard 
interaction vertices, except    in very special kinematical situations.  
Generically, the vertices contain additional modular parameters that must be 
integrated over.     

The paper is organized as follows. In Secs. 2 and 3 we extend the analysis
of Refs. \refs\rtexas\refs\rtext\ to more general diagrams in NCFT at finite temperature
and try to cast the loop amplitudes in a `dual channel' picture,
in terms of tree-level exchanges.   
 In Sec. 4 we will obtain these amplitudes by studying the low
energy SW limit from string theory in order to identify the low-energy winding modes
with undecoupled winding strings. Finally in Sec. 5 we will summarize our
conclusions.

\newsec{Winding modes in noncommutative quantum field theory: an elementary 
example}

The simplest situation where one can formally identify `winding modes'
is that of the two-loop contribution to the free energy in a $\phi^4$
theory. The planar diagram is independent of the deformation parameter 
$\theta^{\mu\nu}$, but a non-trivial phase $\theta(p,q) =  p_{\mu} 
\theta^{\mu\nu} q_\nu$
enters in the loop integral in the non-planar case\foot{We will assume 
throughout that $\theta^{0i}=0$.}
\eqn\nptl{
{\CF}_{\rm NP} = -g^2 \sumint_p \sumint_q {e^{i\theta({\bf p},{\bf q})} 
\over (p^2 +M^2)(q^2 + M^2)},
} 
where we have used the notation
$$
p^2= {\bf p}^2 + {4\pi^2 n^2 \over \beta^2}, \hskip1cm  
\sumint_p \equiv {1\over \beta} \sum_{n\in{\bf Z}}\int {d {\bf p} \over (2\pi)^{d-1}}.
$$
The ultraviolet divergences of this integral can be appropriately eliminated by
renormalization of the $T=0$ limit, as usual in thermal field 
theory \refs\rkbo. In
fact, the ultraviolet structure of this diagram is milder than that of the planar
counterpart, because the divergence contributed by one of the loops is
effectively cut-off by the noncommutative phase provided $(\theta p)^2 \equiv
(\theta^{\mu\nu} p_{\nu})^2$
is non-vanishing. This is an example of the UV/IR mixing of \refs\rp, namely this divergence
will reappear disguised as an infrared effect as $\theta p \rightarrow 0$. 

\fig{\sl 'Channel duality' in the nonplanar self-energy diagram in $\phi^4$ 
theory.}{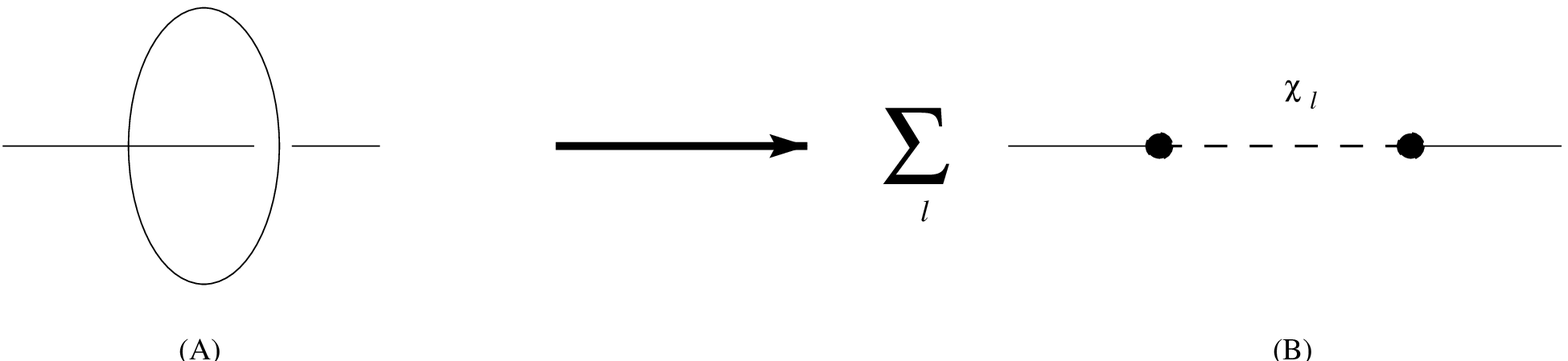}{5truein}

In Ref. \refs\rtexas\ it was pointed out that one could `integrate out' one
of the loops and replace it by a statistical sum over objects living
a the formally T-dual temperature $1/(\theta T)$, thus representing analogues
of winding modes. The essential phenomenon can be understood by simply
looking at the one-loop self-energy tadpole diagram (Fig. 1A)
\eqn\sen{
\Pi (\beta, {\bf p})_{\rm NP} = -g^2 \sumint_q {e^{i\theta({\bf p},{\bf q})} 
\over q^2 + M^2}.
} 
Introducing a Schwinger-parameter representation of the propagator, 
\eqn\sch{
\Pi(\beta, {\bf p})_{\rm NP} = -g^2 \int_0^\infty dt\, \sumint_q e^{-t\left(q^2 + M^2          
-i{\theta({\bf p},{\bf q})\over t}\right)} }
we
can perform the gaussian integral over ${\bf q}$. After a further
Poisson resummation in the thermal frequency running in the loop we obtain
\eqn\refdu{
\Pi (\beta, {\bf p})_{\rm NP} = -{g^2 \over 4 \pi^{d \over 2}} \int_0^\infty ds\,
s^{d-4 \over 2} \,\sum_{\ell\in {\bf Z}} \;e^{-s[\beta^2 \ell^2 + (\theta {\bf p})^2] - 
{M^2\over 4s}},
}   
where we have changed variables to the `dual' Schwinger parameter $s=1/4t$.    
This form is very convenient to perform the subtraction of the $T=0$
 self-energy, since we simply have to restrict the integer sum to $\ell \neq 0$.

For $d<4$, the explicit power of $s^{d-4 \over 2}$ in the proper time integral
 can be `integrated in'
into the exponent by introducing $4-d$ extra gaussian variables ${\bf z}_\perp$
 and we can
write the full non-planar loop in the following suggestive form
\eqn\propchi{
  \Pi(\beta, {\bf p})_{\rm NP} =- \sum_{\ell\in {\bf Z}} \int d{\bf z}_{\perp}
 {|g_{\phi\chi}(\ell,
{\bf p}, {\bf z}_\perp)|^2 \over
 \beta^2 \ell^2 + (\theta {\bf p})^2 + {\bf z}_\perp^2 }.
   }
That is, we have written the original loop diagram in a `dual channel'  
in terms of an infinite number of tree-level exchanges of particles
$\chi_{\ell}$ with momenta $\theta {\bf p}$, mass proportional to $|\beta \ell|$
and extra momentum variables in $d_\chi^\perp = 4-d$ transverse dimensions
 (Fig. 1B). This
complete expression renormalizes the mass of the particle running in the
second loop\foot{One could proceed in the standard way and perform a
resummation of ring diagrams.}. 

The mass of the $\chi_\ell$-fields, 
 scaling as integer multiples of the thermal length, is characteristic 
of winding modes of closed strings. 
The effective coupling squared of
 these particles to the fields in external legs
is given by
\eqn\effcou{
|g_{\chi\phi} (\ell, {\bf p}, {\bf z}_\perp)|^2 = {g^2 \over 4\pi^2} \,\int_{0}^{\infty} 
ds \; e^{-s\,- {1\over 4s}{M^2}[\beta^2 \ell^2 + (\theta {\bf p})^2 + {\bf z}_\perp^2]}.
}
Thus, if the original field was massive, the coupling to the
 $\chi_{\ell}$-field is
suppressed at high values of momentum and winding number $\ell$, i.e. only
fields with winding numbers $|\ell| < (\beta M)^{-1}$ contribute significantly
to the tree-level exchange. The most interesting case is that of a massless
field theory. In this case the effective coupling is constant and weights
all winding numbers democratically, with the coupling strength $g_{\phi
\chi} = g/(2\pi)$. Finally, if the $\phi^4$-field 
is tachyonic, the whole expression is
meaningless, since it diverges at the $s=0$ end. For this matter this
`channel duality' in NCFT is reminiscent of
open/closed-string channel duality in string theory. Since NCFT can be
obtained in many cases as low-energy limits of open-string theories, we
find it natural that   the `dual channel', obtained through a modular
transformation $t=1/4s$, exhibits the open-string tachyon as an ultraviolet
divergence.  
  
It is most interesting to compare the winding $\chi_\ell$-fields we have
defined with the $\psi$-particles of \refs\rp\refs\rrase. The structure of the propagator
shows that these fields are formally similar
\eqn\fs{
\left\langle \chi_\ell ( -{\bf p}, -{\bf z}_\perp) \;\chi_\ell ( {\bf p},
{\bf z}_\perp) \right\rangle
 = {1\over \beta^2 \ell^2 + (\theta {\bf p})^2 + {\bf z}_\perp^2},
 }
namely, they have a `static' kinetic term with the kernel
 $-\partial\circ\partial
 = (\theta
{\bf p})^2$ for a field non-canonical dimension. Furthermore,  at least as long as  
$d\leq 4$, the non-standard power of the propagator can be understood in
terms of a free propagation in a $4-d$ dimensional `transverse bulk'. 
The effective mass $ |\beta \ell |$ plays also the role of the
inverse ultraviolet cutoff $\Lambda^{-1}$ in the treatment of \refs\rp\ and,
 in the
absence of the explicit ultraviolet cutoff, the original
ultraviolet divergence is back as an infrared divergence at $\theta {\bf p} \rightarrow 0$. In
our expression, this shows up as a pole in the zero-winding sector. It is
precisely
this contribution that is subtracted when renormalizing the self-energy by
the zero-temperature one\foot{Notice that this procedure is different 
from the one followed in \refs\rtexas\ where the authors worked with the two-loop
free energy for $\phi^4$ NCFT before subtracting the zero temperature counterterms. 
In that case the ultraviolet divergence in one of the original loops partially 
transforms into an 
infrared one after Poisson resummation and integration over the loop momenta.
This is just a consequence of UV/IR mixing.}. Therefore, 
we confirm that $|\beta \ell|$ plays the role of a regulator.         

One important difference between our tree-level exchange interactions and
the $\psi$-fields of \refs\rp\refs\rrase\
 is that our ultraviolet cutoff, $T$, has a physical
 interpretation, and we are free from the arbitrariness of the choice of
Wilsonian cutoffs. In particular 
we can integrate out the complete non-planar loop in terms
of the infinite tower of tree exchanges of $\chi_\ell$ particles. The
manipulation is not {\it a priori} restricted to the extreme ultraviolet part.  

One interesting aspect of the tree-exchange `dual' representation \propchi\ 
is that it admits an interpretation for the planar diagram too. The
only difference in the planar case comes from setting $\theta {\bf p} =0$.
Therefore, the planar thermal loop can be replaced in this case by
\eqn\proppl{
\Pi(\beta, {\bf p})_{\rm P} - \Pi(\infty, {\bf p})_{\rm P} = - \sum_{\ell \neq 0}
\;\int d{\bf z}_\perp \;
{|g_{\chi\phi} (\ell, {\bf p}=0, {\bf z}_\perp)|^2 \over
\beta^2 \ell^2 + {\bf z}_\perp^2 }.}
Now we must work with the fully renormalized quantity ($\ell \neq 0$ in
the winding sum) and the propagator of the $\chi_{\ell}$ particles is inserted
formally at zero noncommutative
 momentum, i.e. we have a sum over  zero-momentum tadpoles
of the $\chi_\ell$-fields. This is also reminiscent of the closed-string
 interpretation, because closed strings have  tree-level tadpoles on 
D-branes.  
 Now we would be inclined to interpret the residue
of the propagator poles as the product of the couplings  
$g_{\chi \phi \phi} \cdot g_{\chi{\rm\hbox{-}vac}}$ (Fig. 2).  

\fig{\sl Dual channel interpretation of the thermal loop in the planar contribution to the 
two-point function in $\phi^4$ noncommutative field theory.}{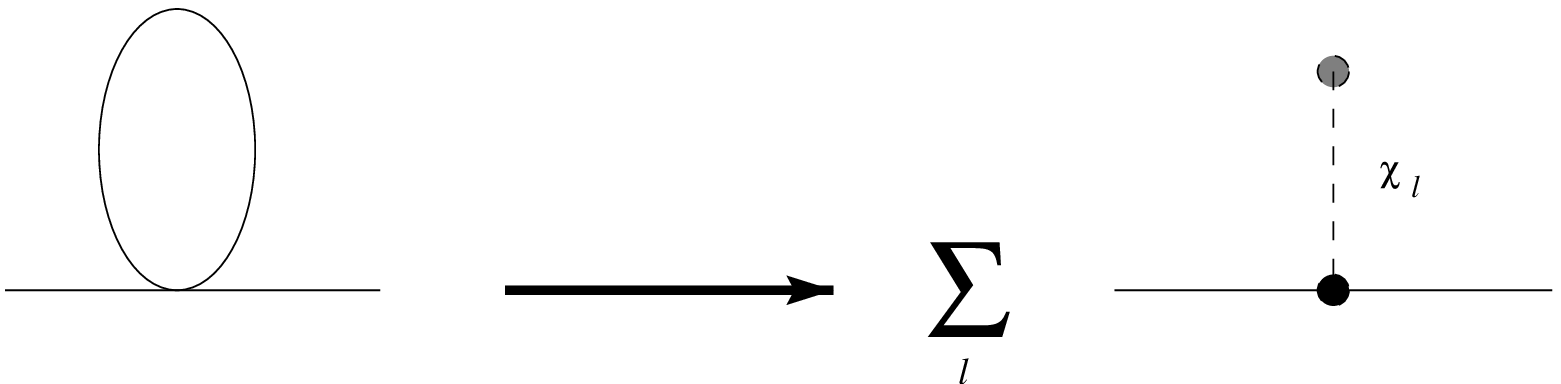}{4truein}

\newsec{Integrating out a general loop} 

In the above example we have seen how the effect of a thermal tadpole loop
in a noncommutative $\phi^{4}$ theory admits a `dual channel' interpretation 
in terms of a the tree-level 
interchange of some `winding' $\chi_{\ell}$-field with inverse propagator 
$\beta^2 \ell^2+(\theta {\bf p})^2$ which mixes with the fundamental $\phi$-quantum.
It would be interesting to decide to what extent this duality between
thermal loops and tree-level $\chi_{\ell}$-exchanges is a general feature of
noncommutative field theory at finite temperature, or just a property of 
a particular class of diagrams and theories.

\subsec{Generic one-loop diagram in noncommutative $\phi^{n}$ theory}

The first case we can consider is the generalization of the example 
studied in the previous section, a generic one-loop diagram with $N=N_{+}+N_{-}$ 
vertices in the noncommutative version of $\phi^{n}$ field theory in 
$d$ dimensions (Fig. 3), where we will take $N_{-}$ vertices as `twisted', so the
amplitude will be nonplanar whenever $N_{\pm}\neq N$. 
Thus, the fully amputated amplitude can be written
as
$$
\CA(p_{1},\ldots,p_{N})= g^{2N}\sumint_{q} \prod_{a=1}^{N}
{e^{-{i\over 2}\xi_{a}p_{a}\theta(q+Q_{a})}\over (q+Q_{a})^2+M^2}\delta(Q_{N}),
$$
where $Q_{a}=\sum_{i=1}^{a}p_{i}$ and $\xi_{a}=\mp 1$ depending on whether the 
insertion is  twisted or not; $p_{a}$ indicates the total 
momentum entering in the loop through the $a$-th insertion. 
Using Feynman and Schwinger parameters we can write
\eqn\scop{
\eqalign{
\CA(p_{1},\ldots,p_{a})&=g^{2N}(N-1)!\,e^{-{i\over 2}\sum_{a}\xi_{a}{\bf p}_{a}\theta 
{\bf Q}_{a}}
\int_{0}^{\infty}dt\,t^{N-1}\, e^{-tM^2}
\int_{0}^{1} [dx] \, e^{-t\sum_{a}x_{a}Q_{a}^2}\cr
&\times {1\over \beta}
\sum_{n\in{\bf Z}}e^{-t\left({4\pi^2n^2\over\beta^2}+{4\pi n\over \beta}\sum_{a}
x_{a}Q^{0}_{a}\right)}
\int {d{\bf q}\over (2\pi)^{d-1}}e^{-t({\bf q}^2+2{\bf q}\cdot \sum_{a}
x_{a}{\bf Q}_{a})}e^{i{\bf p}_{\rm np}\theta{\bf q}},
}
}
the integration measure $[dx]$ over the Feynman parameters $x_{a}$ ($a=1,\ldots,N$) 
is given by
$$
[dx]\equiv \delta\left(\sum_{a=1}^{N}x_{a}-1\right)\prod_{a=1}^{N}dx_{a}
$$
and ${\bf p}_{\rm np}$ denotes the total {\it nonplanar} spatial momentum entering in the 
loop through the
$N_{-}$ `twisted' insertions,
${\bf p}_{\rm np}\equiv -\half \sum_{a=1}^{N}\xi_{a}
{\bf p}_{a}$. 

By integrating the loop 
spatial momentum and performing a Poisson resummation the total amplitude can 
be recast in terms of the dual Schwinger parameter $s=1/(4t)$ in the form
\eqn\recall{
\eqalign{
\CA(p_{1},\ldots,p_{N})&=g^{2N}{(N-1)!\over 2^{2N}\pi^{{d\over 2}}}\,\CW_{\rm NC}
 \cr
&\times \int_{0}^{\infty}ds\, s^{d-2N-2\over 2}\sum_{\ell\in{\bf Z}}
e^{-s\left[\beta^2\ell^{2}+(\theta {\bf p}_{\rm np})^2\right]} F_\ell
(s;\beta,p_{1},\ldots,p_{N}),
}
} 
where $\CW_{\rm NC}$ is the overall noncommutative  phase of the diagram
\eqn\fas{
\CW_{\rm NC}=e^{-{i\over 2}\sum_{a=1}^{N}\xi_{a}{\bf p}_{a}\cdot \theta{\bf Q}_{a}}
}
and the function $F_\ell(s;\beta,p_{a})$ is expressed in terms of an integral over the
$x^{a}$ as
\eqn\F{
F_\ell (s;\beta,p_{a})=e^{-{1\over 4s}M^2}\int_{0}^{1} [dx] 
e^{-{1\over 4s}\left[\sum_{a}x_{a}Q_{a}^{2}
-\left(\sum_{a}x_{a}Q_{a}\right)^2\right]}e^{i\beta\ell
\sum_{a}x_{a}(Q_{a}^{0})^2}
e^{i\sum_{a}x_{a}{\bf Q}_{a}\cdot(\theta {\bf p}_{\rm np})}.
}

As in the simpler case of the tadpole of the $\phi^{4}$ theory, whenever $d<2N+2$ we can 
replace the factor $s^{d-2N-2\over 2}$ by an integral over $2+2N-d$
 extra variables, 
so we can finally write the diagram in the form of a tree-level exchange of
effective fields propagating in $d_\chi^\perp = 2N +2 - d$ additional `bulk'
dimensions, 
\eqn\genlo{
\CA(p_{1},\ldots,p_{N})=\sum_{\ell\in {\bf Z}}\int d{\bf z}_{\perp}{f(\ell,p_{a},{\bf z}_{\perp})
\over\ell^{2}\beta^2+(\theta {\bf p}_{\rm np})^2+{\bf z}_{\perp}^{2}}
}
where the function $f(p_{a},\ell)$ is given by
$$
f(\ell,p_{a},{\bf z}_{\perp})=
g^{2N}{(N-1)!\over 2^{2N}\pi^{d\over 2}}\, {\cal W}_{\rm NC} 
\;\int_{0}^{\infty}ds\,
F_\ell \left[{s\over \ell^2\beta^2+(\theta{\bf p}_{\rm np})^2
+{\bf z}_{\perp}^{2}};\beta,p_{a}\right].
$$ 

\fig{\sl Channel duality for a nonplanar thermal loop in $\phi^n$ noncommutative 
field theory.}{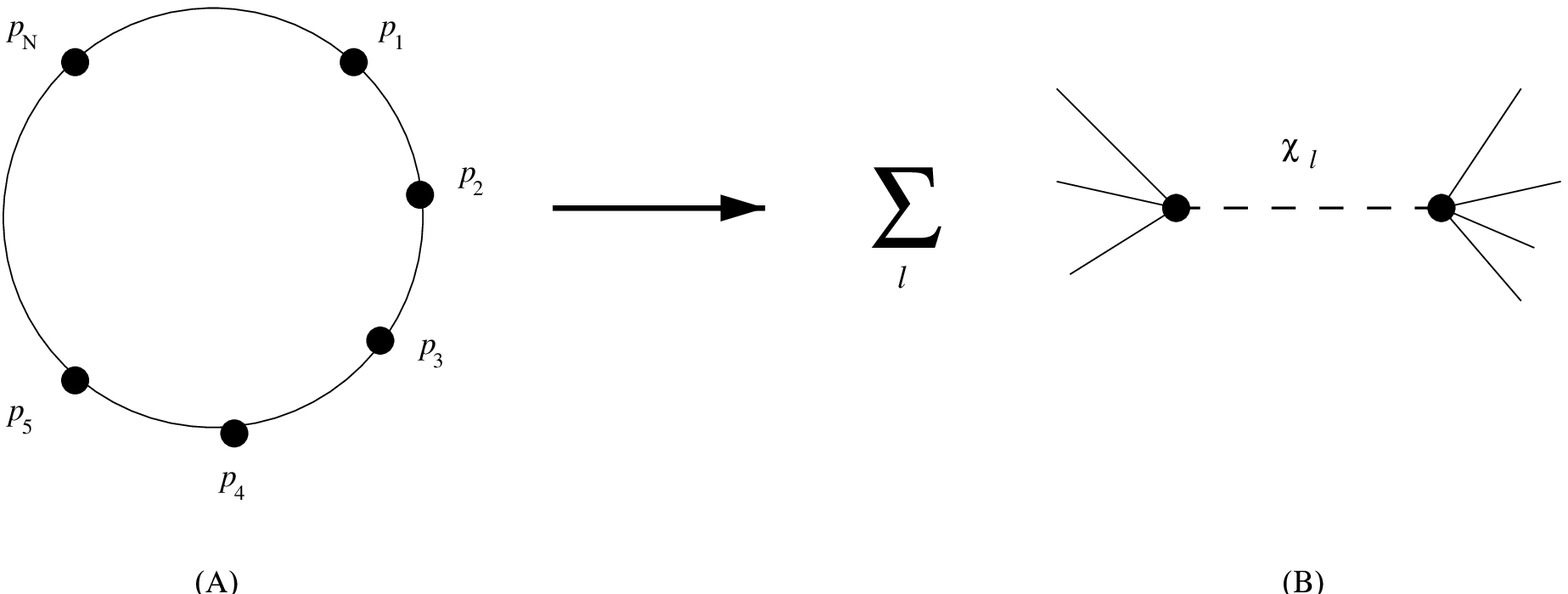}{5truein}

In the same spirit of the $\phi^4$ tadpole one would like to interpret the 
amplitude \genlo\ as a `dual channel' representation of the original loop diagram 
in terms of a tree-level exchange of $\chi_{\ell}$-particles
with propagators \fs, so
the function $f(\ell,p_{a},{\bf z}_{\perp})$ would be interpreted as the product of 
the couplings in Fig. 3B 
\eqn\glco{
g_{(\phi^{nN_{+}})\chi} \;g_{(\phi^{nN_{-}})\chi} \sim f(\ell,p_{a},{\bf z}_{\perp}).
}
However, such an identification is rather problematic.
 Unlike the case of the $\phi^4$ tadpole, 
there seems to be no unambiguous way to define the individual couplings
$g_{(\phi^{nN_{+}})\chi}$ and $g_{(\phi^{nN_{-}})\chi}$, since their product \glco\
is expressed in terms of a function which does not factorize into the contributions
of the two vertices. Moreover, because of the integration over the Feynman
parameters in Eq. \F, the interaction 
on the two vertices cannot be disentangled, even for massless fields.
It is only in the tadpole case ($N=1$) that
 the integration over Feynman parameters
 disappears
and the whole loop can be understood as resulting from the mixing of the
field $\phi$ with an effective $\chi$-field, thus generalizing the result of the previous
section to $\phi^{n}$.

Therefore, even if we can formally replace the generic thermal loop by the exchange of an effective
$\chi_{\ell}$-particle, we cannot assign ordinary Feynman rules to this field, since 
the total amplitude is expressed as a convolution of the two interaction 
vertices, and
not just a product as it is the case of ordinary
 (and noncommutative) quantum field theory. We summarize this state of
affairs by saying that the vertices of the $\chi_\ell$-fields have
relative moduli that must be integrated over.   

In principle, the $\chi_\ell$-fields introduced
 here could become {\it bona fide}
fields, with standard Feynman rules,
 when considering only the  behaviour of the   diagram at singularities
of the integral over Feynman parameters.  
We suspect that this is the precise
link between the $\chi_\ell$-fields defined here and the $\psi$-fields
of ref. \refs\rp\refs\rrase.     

 The appearance of moduli in the `dual channel vertices' will find
a string-theory explanation in the next section. First, we shall discuss
some special instances in which the formalism simplifies.   

\subsec{Some special cases at  two loops}

The previous example seems to indicate that, although in general we can replace 
nonplanar loops in NCFT by tree-level exchanges of an infinite tower of
 some effective $\chi_\ell$-fields, 
in a generic situation the nonlocal character of this field makes the effective
description not very transparent. Here we will further comment on two examples
where 
this effective description is useful. 

Let us first consider noncommutative super Yang--Mills (NCSYM) theories at finite temperature. 
The two loop free energy density can be written for $U(N)$  NCSYM$_{d}$
as \refs\ravm
\eqn\sym{
\eqalign{
\CF(\beta,\theta)=& \CF(\beta,\theta=0)+ \CC_{\rm sc}\,g^2 N \left\{
\int{d{\bf p}\over(2\pi)^{d-1}}\left[{n_{b}({\bf p})\over 
\omega_{p}}+{n_{f}({\bf p})\over \omega_{p}}\right]^{2} \right.\cr
 -& \left. \int{d{\bf p}\over (2\pi)^{d-1}}\int{d{\bf q}\over (2\pi)^{d-1}}
\left[{n_{b}({\bf p})\over \omega_{p}}+{n_{f}({\bf p})\over \omega_{p}}\right]
\left[{n_{b}({\bf q})\over \omega_{q}}+{n_{f}({\bf q})\over \omega_{q}}\right]
e^{i\theta({\bf p},{\bf q})}
\right\},
}
}
where $\CC_{\rm sc}=16,4,1$ for theories with 16, 8 and 4 supercharges respectively  
\refs\vm, $\omega_{p}=|{\bf p}|$ and $n_{b(f)}({\bf p})=(e^{\beta|{\bf p}|}\mp 1)^{-1}$ 
are the Bose--Einstein and Fermi--Dirac distribution functions. It is interesting to notice
how, for 
NCSYM theories, 
the `nonplanar' part of the two-loop free energy [the last term in \sym] factorizes 
into the product of two independent loop contributions only linked through the 
noncommutative phase, 
much in the same fashion of $\phi^{4}$ NCFT. Following Ref. \refs\rtexas\ we can now 
integrate one of these loops to try to spot winding states (Fig. 4). When $d<4$ again we can
introduce $4-d$ extra variables ${\bf z}_{\perp}$ to write
\eqn\impares{ 
\eqalign{
\int{d{\bf q}\over (2\pi)^{d-1}}\left[{n_{b}({\bf q})\over 
\omega_{q}}+{n_{f}({\bf q})\over \omega_{q}}\right]e^{i\theta({\bf p},{\bf q})}
=&{1  \over (2\pi)^2}\sum_{\ell\in{\bf Z}}
\int d{\bf z}_{\perp}{1 + (-1)^{\ell +1} 
\over \ell^2\beta^2+(\theta{\bf p})^2+{\bf z}_{\perp}^
2} \cr 
=&{1\over 2\pi^2}\sum_{\ell\in{\bf Z}}
\int d{\bf z}_{\perp}{1\over (2\ell+1)^2\beta^2+(\theta{\bf p})^2+{\bf z}_{\perp}^2}
}
}
\fig{\sl Dual channel description of the  nonplanar thermal loop in $\phi^n$-like 
two-loop nonplanar vacuum diagram.}{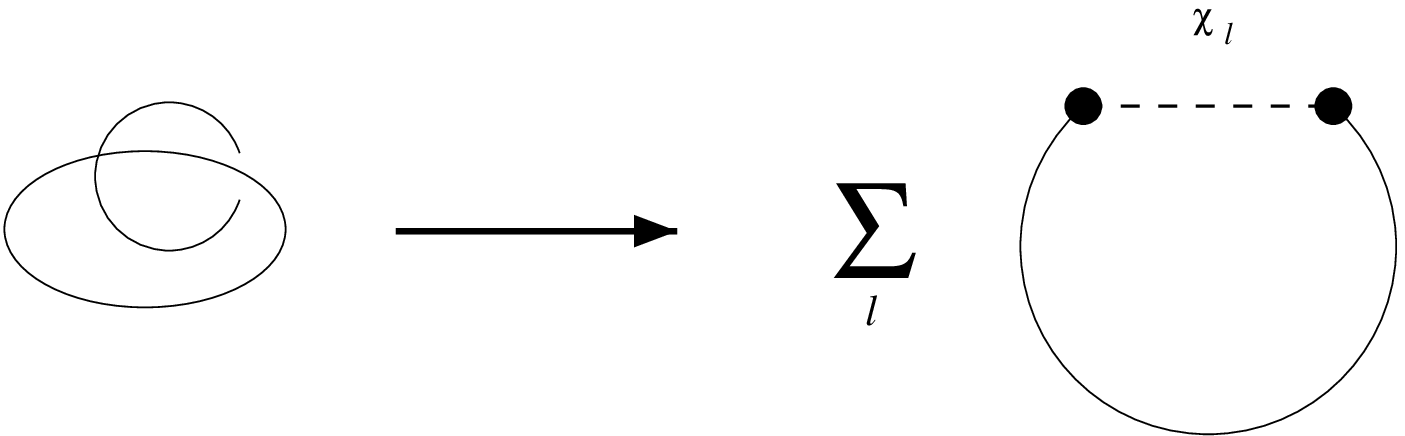}{4truein}
That is, we find a standard tower of $\chi_\ell$ particles, restricted
to {\it odd} winding numbers.  From this expression we learn that,  
in general, the fermion loops will give rise to $\chi_\ell$ particles with
negative norm for even $\ell$. For the supersymmetric case, there is
a cancellation with the  tower coming from the bosonic loop
 and we find the projection onto odd
winding numbers. We shall  give a string theory explanation of this phenomenon
in the next section.

Notice that the factorization of the two-loop free energy into contribution
of two independent loops is not a general property of any quantum field theory. 
In particular,
for $\phi^3$ NCFT the integrand of  $\CF(\beta,\theta)$ does
not have this property even in the massless case. What is special about 
NCSYM is the fact that many of the individual diagrams 
contributing to  \sym\ have loops with two external insertions, so
one would    need at least one Feynman parameter in order
to formally `integrate out' the loop, along the lines of the general
discussion above. Yet, the complete two-loop diagram shows factorized
form and one can introduce effective $\chi_\ell$ particles with
standard Feynman rules (apart from the negative-norm feature in the
fermionic case). 

The reason behind this simplification is two-fold. First, gauge symmetry
relates the $\phi^3$-like diagrams to the $\phi^4$-like diagrams. Second,
the theory is massless, so that the effective coupling \effcou\ is a 
momentum-independent pure
number, and  thus  both $\phi^4$-like loops are  completely disentangled
from the kinematical point of view. Therefore, this factorization is, in principle,
specific of two-loop diagrams in massless theories whose symmetries can relate
all diagrams, contributing to a given physical 
quantity,  to $\phi^4$-like ones.    Another example, considered
in \refs\rtexas, is the   massless Wess--Zumino model, where supersymmetry
plays the relevant role. In the massless limit the Wess--Zumino model 
reduces itself to a supersymmetric version of $\phi^{4}$ NCFT. Thus the
factorization of the two-loop free energy follows from the
factorization of the corresponding diagram in $\phi^{4}$ theory under
the substitution $n_{b}({\bf p})\rightarrow n_{b}({\bf p})+n_{f}({\bf p})$. 
In the NCSYM case it is not supersymmetry, but
rather gauge symmetry, the one playing the simplifying role, because
the two-loop factorization is true already for nonsupersymmetric 
noncommutative Yang--Mills theories \refs\ravm.    

\newsec{Windings and closed strings}

\subsec{Heuristic considerations}

Given that many NCFT derive from open-string theory in background
$B$-fields in the SW limit, it is natural to associate the winding
modes of the previous representations to closed strings in 
intermediate states. Namely, the structure of \propchi\ is reminiscent
of a closed-string tree-level propagator between 
boundary states of a D$_{d-1}$ brane (Fig. 5). 
Heuristically, we expect 
\eqn\heu{
 \Pi(\beta, p)_{\rm NP} =- \sum_{\ell} \int d{\bf z}_{\perp}
 {|g_{\chi}(\ell,
p, {\bf z}_\perp)|^2 \over
 \beta^2 \ell^2 + (\theta p)^2 + {\bf z}_\perp^2 }
\sim \lim_{\rm SW} \;\left\langle {\rm D}_{d-1}; V_p {\Big |} {1\over 
\Delta_{cl} } {\Big |} \;{\rm D}_{d-1}; V_p \right\rangle,
} 
whereas the planar diagram would be a low-energy limit of $\left\langle 
{\rm D}_{d-1}; V_p, V_{-p} | \Delta_{cl}^{-1} | {\rm D}_{d-1} \right\rangle$, for
suitably defined boundary states.
\fig{\sl Seiberg--Witten limit of the nonplanar two-point 
function in the closed-string channel.}{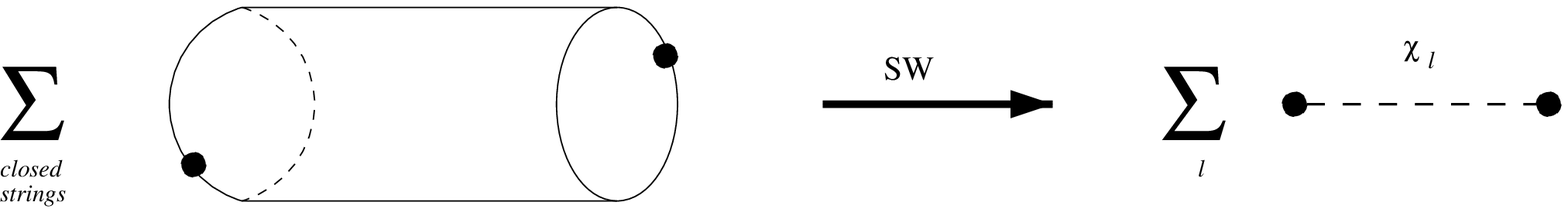}{6truein}

There are various pieces of the previous tentative equation that fit
nicely. First, the closed-string inverse propagator is
\eqn\iprop{
\Delta_{cl} = {\alpha' \over 2} \left(g^{\mu\nu} p_{\mu} p_{\nu} + 
{\beta^2 \ell^2 \over 4\pi^2 \alpha'^2} + M_{cl}^2 \right)}
where $g_{\mu\nu}$ is the closed-string or sigma-model
 metric, to be distinguished
from $G_{\mu\nu}$ or open-string metric. The precise
relation is defined in \refs\rsw:
\eqn\defmet{
G^{\mu\nu} = \left({1\over g + 2\pi\alpha' B}\right)_S^{\mu\nu} ,\qquad
\theta^{\mu\nu} = 2\pi\alpha' \,
\left({1\over g + 2\pi\alpha' B}\right)_A^{\mu\nu} 
}
where by the subscripts $S$ and $A$ we indicate the symmetric and antisymmetric
part respectively. We can take
$g_{\mu\nu} = \delta_{\mu\nu}$ in commutative directions, including the
$d_\perp = 
D-d$ Dirichlet--Dirichlet directions transverse to the ${\rm D}_{d-1}$
 brane. On
the other hand, in the noncommutative directions the SW scaling assigns
\eqn\swmet{
g^{\mu\nu} \rightarrow -{1\over 4\pi^2 \alpha'^2} (\theta^2 )^{\mu\nu}  
}
as $\alpha' \rightarrow 0$, with  $G_{\mu\nu} = \delta_{\mu\nu}$ and 
 $\theta^{\mu\nu}$ fixed.
 Therefore, the inverse propagator scales                       
\eqn\nprop{
\Delta_{cl} = {1\over 8\pi^2 \alpha'} \left[ \beta^2 \ell^2 + (\theta p)^2 
+ \alpha'^2 {\bf p}_\perp^2 + \alpha'^2 M_{cl}^2 \right]}
and we see that our  familiar combination $\beta^2 \ell^2 + (\theta p)^2$ 
scales together and {\it dominates} over the other terms in the SW limit, since
$\alpha'^2 M^2_{cl} \sim \alpha' N_{\rm osc} \rightarrow 0$.
This is the main evidence for the stringy origin of the winding modes. Indeed,
we have Neumann boundary conditions in the thermal circle, and therefore
the closed-string cylinder can wind in this direction. On the other hand,
there cannot be momentum flow through the Neumann directions unless it is
explicitly inserted via the open-string vertices into the boundary states,
but there is an arbitrary flow of momentum in Dirichlet--Dirichlet directions.
In the noncommutative directions, having a nonzero $B$-field, 
 one could have momentum flow induced just by
the $B$-field. The boundary conditions 
set to zero only  a linear combination of momentum and winding numbers.  
  However, since we are assuming a noncompact D-brane in
spatial directions, there are no winding modes in spatial directions and
thus no extra momentum flow induced by the $B$-field.

If we are willing to naively neglect the nominally subleading terms in
\nprop\ we can almost get  \heu\   with the coupling of the
 $\chi_{\ell}$-fields 
 defined through the `sum rule' over all
 closed-string fields $|\Psi \ket$ (the oscillator excitations)
\eqn\fsr{
|g_{\chi \phi}|^2 \rightarrow
  \sum_\Psi \left\langle {\rm D}_{d-1}; V_p | \Psi \right\rangle
\left\langle \Psi | \,{\rm D}_{d-1} ; V_p \right\rangle }
in the SW limit, perhaps with appropriate powers of $\alpha'$ in front.
According to this picture, the low-energy $\chi_\ell$-fields {\it are not}
some low-lying
  closed-string modes that fail to decouple. In fact, the whole infinite
tower of closed-string modes fails to decouple, but the interaction
with the boundary states defines  an effective coupling for
the $\chi_{\ell}$-field, which represents the coherent exchange of an infinite
number of closed-string excitations.
 Formally, the SW limit squeezes the complete tower of string excited states
 into an approximately continuous band, as compared to the gap of
the winding modes  
\eqn\gapw{
{{\rm Oscillator \;\; Gap} \over {\rm Winding \;\;Gap
} } \sim {\alpha' \over \beta^2} \rightarrow 0.}
This is an interesting compromise between the general lore that 
the closed-string channel should be intractable whenever the open-string
channel is simple \refs\rjaume, and the factual existence of the
dual channel representation in terms of the   $\chi_{\ell}$-fields.

Actually, the sum rule \fsr\ is too naive. 
The first indication that something is missing in \fsr\  is the fact
that a naive attempt to associate the ${\bf z}_\perp$ degrees of
freedom with Dirichlet--Dirichlet momenta ${\bf p}_\perp$ 
in the D-brane codimension
fails quantitatively, because in general $d_\perp \neq d_\chi^\perp= 
2+2N-d$ for a
 ${\rm D}_{d-1}$ 
 brane.
The resolution of the puzzle amounts to recognize that one cannot
simply neglect $\alpha'^2 M_{cl}^2$ in the closed-string propagator, as
compared to $\beta^2 \ell^2$, even in the low-energy SW limit, because
there are an infinite number of states contributing to the sum. In other words,
the truncated sum rule \fsr\ is not convergent in general.

The second reason for concern lies in the definition of the low-energy
effective coupling $g_{\chi \phi}$ as a proper effective vertex. In the
full string theory diagram, the boundary states with open-string insertions
have  moduli (the Koba--Nielsen parameters) that must be integrated over.
Therefore, the stringy diagram does not have in general the structure of
an ordinary tree-level exchange, since both vertices are convoluted in an integral
over Koba--Nielsen parameters. Only at the boundaries of the moduli space,
when the string diagram degenerates into proper field theory diagrams,
one finds standard Feynman rules. This means that the correct sum rule
replacing \fsr\ must hold for the integrand over moduli space (including
the Koba--Nielsen moduli).  

\subsec{Open-string channel}

In the remainder of this section we obtain the correct sum rule by 
careful consideration of the general one-loop open-string diagram.  
This is a weighted sum over spin structures $\{\sigma\}$, each one given
by a path integral on the annulus with 
 arbitrary vertex operator insertions on both the
inner ($-$) and outer (+) boundaries
\eqn\ampl{
\eqalign{ {\cal A} = \sum_\sigma C_\sigma {\cal A}_\sigma =&  
\sum_\sigma C_\sigma\,
  \int_0^\infty {d\tau \over 2\tau} \int_0^\tau [dy^\pm] 
\left\langle \prod_{y^\pm} V_\phi (p^\pm, y^\pm)
\right\rangle_\sigma \cr  =& (\alpha')^{\CN_0} \; G_s^{N/2}   
\;\sum_\sigma C_\sigma \,\int_0^\infty {d\tau \over 2\tau} 
Z(\tau)_\sigma \int_0^\tau [dy^\pm] \;{\cal V}(p^\pm ,y^\pm, \tau)_\sigma ,
}}
where $\tau$ is the modulus of the annulus, 
$y^{\pm}$ are the Koba--Nielsen parameters in each boundary, ${\cal V}(
p^\pm ,y^{\pm}, \tau)$ is the normalized correlator of vertex operators in the 
spin structure $\sigma$, and $Z(\tau)_\sigma$ is the normalization, i.e. 
the path integral without vertex insertions. The power of $G_s$ comes from
the normalization of the vertex operators with the open-string coupling
defined in \refs\rsw
\eqn\defgs{
G_s = g_s \,\left[{{\rm det}\,G \over {\rm det}\,(g+2\pi\alpha' B)} \right]^{
1\over 2} = {(\alpha')^{4-d \over 2} \over (2\pi)^{d-3}} \;g_{\rm YM}^2.
}
The complete amplitude has an appropriate power of the Regge slope $\CN_{0}$
so that the SW limit with fixed $g_{\rm YM}$ 
 leads to the field-theoretical expression of the
amplitude. This depends on the overall dimension of the
amplitude and the number of insertions.

The function ${\cal V}(p^\pm ,y^{\pm}, \tau)$ 
 is a polynomial in external field polarizations, the fermionic Green function $\CG_{\rm F}$
and  derivatives of the bosonic Green function
${\cal G}_{\rm B}$
  on the annulus with appropriate $B$-dependent boundary conditions, times
the contraction of the tachyonic part of the vertices
\eqn\corr
{{\cal V}(p^\pm,  y^\pm, \tau )  = {\cal P}(\phi, p^\pm ,\CG_{\rm F}, 
 \partial {\cal G}_{\rm B}) \,e^{ p 
\cdot {\cal G}_{\rm B} \cdot p}.}  
This  Green function can be parametrized completely by the open-string
metric $G^{\mu\nu}$ and noncommutativity parameters $\theta^{\mu\nu}$, except for a single
constant term from the purely bosonic component \refs\rcallan,
 which contributes to nonplanar diagrams and depends explicitly
on the sigma model metric  
$g_{\mu\nu}$.  If we separate this contribution from \corr\ we   can write 
\eqn\corrr{
{\cal V} (p^{\pm}, y^\pm, \tau) = {\overline {\cal V}}
 (p^\pm, y^\pm, \tau) \;{\rm exp}\left[-{\alpha'
\pi^2 \over \tau} g^{\mu\nu} ( p_\mu p_\nu )_{\rm cyl}\right] }
where $p_{\rm cyl} = \sum p^+ = -\sum p^-$ is the total momentum circulating
in the closed-string channel. We recognize this term as the standard kinetic
term in $\Delta_{cl}$ \iprop.

 If we assume, for simplicity, that
the external insertions are space-time bosons  with  
 no thermal frequency, i.e. we have a purely
static bosonic correlator,
 then the world-sheet partition sum can be written directly
in operator form as
\eqn\vacp{
Z(\tau)_\sigma = {1\over {\rm Vol}_G} \;{\rm Tr}_{\rm open}
 \;{\cal S}_{\sigma} \;e^{-\tau\,\Delta_\sigma},  
}
where ${\cal S}_\sigma$ is a piece of the GSO projector,
$\pm {1\over 2}$ or $\pm {1\over 2}(-1)^F$ depending on the
spin structure,  $\Delta_\sigma$
 is the open-string world-sheet hamiltonian, and we have normalized by
the volume in the open-string metric $G_{\mu\nu}$. The  
temperature-dependent part of \vacp\
 is unaffected by the $B$-field, as long as
we keep $B_{0i} =0$. Therefore, it has the form 
\eqn\opham{
\Delta (\beta)_\sigma =
 \alpha' \,(p^0)^2 = \alpha' {4\pi^2 n_\sigma^2 \over \beta^2}, 
}
with $n_\sigma$ integer in those spin structures running space-time bosons 
in the loop, and half-integer in those running  space-time fermions.
{}From here we can read off the relation between the annulus modular parameter
and the low-energy Schwinger parameter of the field theory expressions in
eq. \scop\ of the previous section; it is
\eqn\modsch{
t=\alpha'\,\tau.
}

For the comparison with the low-energy expression, it is useful to
work with Koba--Nielsen parameters normalized to unity, $y=\tau \,x$, so
that a further factor of $\tau^N = (t/\alpha')^N$
 appears for a total of $N$ insertions. Consistency of the SW
low-energy limit requires that, for an appropriate choice of $\CN_0$, 
  the field-theoretical amplitude is obtained as
\eqn\lowenam{
{\cal A}_{\rm NCFT} =  
\lim_{\rm SW} \; (\alpha')^{{\cal N}_0 + {4-d \over 4} N} \,g_{\rm YM}^N 
\int_0^\infty dt\,t^{N-1} \sum_\sigma C_\sigma \int_0^1 [dx^{\pm}]
 Z\left({t\over \alpha'}\right)_\sigma \,{\cal V}\left(p^{\pm}, x^{\pm}, {t\over \alpha'}
\right)_\sigma.}
The precise details of this limit in various examples of the bosonic theory
at zero temperature can be found in recent papers 
\refs\rad\refs\rkl\refs\rbil\refs\rjaume\refs\raroz\refs\rlmi. 
The important feature  is that the SW limit is dominated
by massless open strings (in the bosonic examples one is forced to 
discard the open-string tachyon by hand). Thus, in comparing with \scop,
we must set $M=0$ and interpret the normalized Koba--Nielsen parameters
$x^\pm = y^\pm /\tau$ as Feynman parameters of the field theory diagram.

 On general grounds, the massless open-string dominance means that
 we do not expect
the closed-string channel expression to be simple, in the sense of being
saturated by a finite number of closed-string fields.

\subsec{The closed-string channel sum rule}

Ideally, we would like to specify explicitly, in the closed-string
Fock space,  the boundary states  appearing in eq.  
\heu. This is a very complicated task in general, and can be 
carried out in detail only for the `vacuum' boundary states
$|{\rm D}_{d-1}\ket$ without open-string vertex insertions. On
the other hand,  we can  
obtain explicitly the overlap in \heu, by direct modular transformation
of the open-string channel expression \ampl.  
 
 In order to keep track of the right normalization of
winding modes, we  perform  a Poisson
resummation of the discrete frequency sums
\eqn\poi{
{1\over \beta} \sum_{n_\sigma} e^{-\tau 
  \Delta(\beta)} =  (4\pi \alpha' \tau)^{-1/2} \,\sum_{\ell\in {\bf Z}}
e^{-{\beta^2 \ell^2 \over 4\alpha' \tau}} \; (-1)^{\ell {\bf F}_\sigma} 
}
where ${\bf F}_\sigma$ is the {\it space-time} fermion number in the
open-string channel. It is correlated with the closed-string sector
in the cylinder channel in such a way 
that ${\bf F}=0$ corresponds to the NS--NS
sector of the D$p$-brane boundary state, whereas ${\bf F}=1$ leads to
the R--R exchange \refs\rvmdb. Therefore, we have found that the winding modes of
closed strings in the R--R sector are weighted by the so-called
Atick--Witten phase $-(-1)^\ell$ \refs\raw,
the extra minus sign coming from the GSO projection, or more elementarily,
from the overall minus sign of fermion loops in space-time. We recognize
in the phase $(-1)^{\ell +1}$ the effect pointed out in eq. \impares. 
Namely, fermionic loops in the `open-string channel' lead to phases
in winding modes in the `closed-string channel'. In particular, in the
supersymmetric case we also find a projection onto odd winding numbers
in the full string theory expression.  The closed-string interpretation
also explains the `negative norm' of the tower of even $\chi_{2\ell}$-fields
coming from a fermion loop; it is just an effect of the D-brane carrying
`axionic' charge with respect to these fields.

 In view
of the closed-string propagator in   
\iprop, the appropriate modular transformation to obtain ${\rm exp}(-\tau_2 \,
\Delta_{cl})$ is $\tau_2 = 2\pi^2 /\tau$. The non-trivial piece of the
overlap \heu\ is that of the oscillator degrees of freedom. We define  
\eqn\modtrans{
Z\left({2\pi^2 \over \tau_2}\right)_\sigma^{\rm osc} 
\,{\overline{\cal V}}
\left(x^\pm, p^\pm,   
{2\pi^2 \over \tau_2}\right)_\sigma  =
 \left({\tau_2\over \pi}\right)^{{2-D \over 2}+N} 
\left\langle         
{\rm D}_{d-1} ; V_{p^+}, x^+ {\Big |} e^{-\tau_2 \Delta_{cl}^{\rm osc} }
{\Big |} {\rm D}_{d-1} ; V_{p^-}, x^- \right\rangle_\sigma }
where 
$$
\Delta_{cl}^{\rm osc} = {\alpha' \over 2} M^2_{cl}.
$$
Numerical constants have been absorbed in the definition of the
boundary states. The modular anomaly, depending on the total 
dimension where the
string oscillates ($D=10$ for superstrings), is captured by looking
at the case without insertions, where an explicit construction of
the boundary states exists.                      
Taking into account the factor of $\tau_2^{d-1 \over 2}$
from the integral over world-volume momenta in the
evaluation of $Z(\tau)_\sigma$,  
 we find the total moduli-space measure to be        
\eqn\modtot{
d\tau_2 \,\tau_2^{-{d_\perp/ 2}} \;[dx^\pm].
}
The factor of $\tau_2^{-d_\perp /2}$ can be `integrated in' by introducing
explicitly the integral over the $(D-d)$-dimensional 
 transverse momenta ${\bf p}_\perp$, which
replaces $M^2_{cl} \rightarrow M_{cl}^2 + {\bf p}_{\perp}^2$ and then we
have a standard measure for a propagator, as in \heu. Notice that the
power of $\tau_2^{N}$ in \modtrans\ is crucial in obtaining \modtot, so
that the only $\tau_2$-dependence of the overlap is in the world-sheet
 evolution  
operator. 
We see explicitly how the proper counting of transverse dimensions, as
read-off from the powers of Schwinger parameters, is working fine thanks
to the modular anomaly in \modtrans. 
 
Collecting all terms, we can write a closed-string channel expression
for the total amplitude in the SW limit. In order to make contact with
the expressions in section 3, we define a dual Schwinger parameter with
mass squared dimension
$$
s= {\tau_2 \over 8\pi^2 \alpha'}   
$$   
in terms of which, we have 
\eqn\cltotal{
\eqalign{{\cal A}_{\rm NCFT}  &= \lim_{\rm SW} \;\; g_{\rm YM}^N \; (\alpha')^{
{\cal N}_0 + {4-d \over 4}\, N} \;
 \int_0^\infty ds \;s^{-d_{\perp} / 2}\int_{0}^{1}[dx^{\pm}]  \cr
&\times \sum_\sigma C_\sigma \sum_{\ell \in {\bf Z}} e^{-s[\beta^2\ell^2 +
(\theta p)^2]} \,U_{\ell,\sigma}                
\left\langle
{\rm D}_{d-1} ; V_{p^+}, x^+ {\Big |} e^{-s (2\pi\alpha' M_{cl})^2 }
{\Big |} {\rm D}_{d-1} ; V_{p^-}, x^- \right\rangle_\sigma  
}
}
with 
$U_{\ell, \sigma}$ is the Atick--Witten phase. 
We can recast this expression in the form of eq. 
\recall 
\eqn\otravez{
 {\cal A}_{\rm NCFT}  = 
 \lim_{\rm SW} \;\CW_{\rm NC}\;g_{\rm YM}^N  \int_0^\infty ds\,
s^{-N +{d-2 \over 2}} \;\sum_{\ell \in {\bf Z}} e^{-s[\beta^2\ell^2 +
(\theta p)^2]} \int_0^1 [dx^\pm]\; F_\ell (s, p^\pm, x^\pm )
} 
where $\CW_{\rm NC}$ is the global noncommutative phase of the diagram given
by Eq. \fas,  up to numerical constants.   This expression implies
 a `sum rule' for the function $F_\ell (s, p, x)$: 
\eqn\truesr{
\eqalign{
\CW_{\rm NC}F_\ell (s, p^\pm , x^\pm ) =& \lim_{\rm SW} \;\; (\alpha')^{
{\cal N}_0 + {4-d \over 4}\, N}  \;s^{{2-D \over 2}+N} \sum_\sigma C_\sigma
 \,U_{\ell,\sigma} \cr
\times & \sum_{\Psi} \left\langle
{\rm D}_{d-1} ; V_{p^+}, x^+ {\Big |} \Psi\right\rangle_\sigma \,
 \left\langle 
\Psi {\Big |}  e^{-s (2\pi\alpha' M_{\Psi})^2 }
{\Big |} \Psi \right\rangle_\sigma \left\langle \Psi {\Big |}
 {\rm D}_{d-1} ; V_{p^-}, x^- \right\rangle_\sigma
} } 
where the sum over closed-string states $|\Psi \ket$ runs over
all oscillator degrees of freedom of the closed strings in the bulk.
In comparing \truesr\ with the field-theoretical expression in \F, 
we must take into account that $M=0$ and \F\ was derived for a purely
bosonic loop, hence there is no non-trivial Atick--Witten phase in
\F. Furthermore, \truesr\ was derived under the assumption that 
external states were bosonic and {\it static}, i.e. external momenta
have  vanishing
time components. That explains the absence of the phase 
$
{\rm exp}[i\beta \ell \sum_a x_a (Q_a^0)^2]
$
in \truesr.

 We  conclude  this subsection with some observations on the interpretation 
of  \truesr:  

{\it i}) In terms of the dimensionless closed-string modulus, 
the SW limit in \truesr\ takes $\tau_2 \rightarrow 0$. In this region
of moduli space, the infinite tower of closed-string fields contributes
to the sum rule, which is by no means saturated by a few closed-string
fields. This was already obvious from the fact that the open-string
channel expression {\it was} saturated by massless open strings. 
 
{\it ii}) The sum rule \truesr\ replaces the naive one in \fsr. One of the defects
of \fsr, the mismatch
  between the powers of the Schwinger parameter and the true number of
transverse dimensions of the brane, is resolved by noticing that the
sum rule includes a non-trivial power of $s^{D-2 \over 2}$, together
with an explicit exponential kernel, which gives back the field-theoretical
measure in \recall\ in the SW limit, and ensures the convergence of
the sum over closed-string states. In fact, since this limit takes 
$\tau_2 \sim \alpha' s \rightarrow 0$, the way to evaluate the infinite
 sum over
states  in \truesr\ is to perform a modular transformation back to the
open-string variables. In this process we get the appropriate powers
of the Schwinger parameter from the modular anomaly of the oscillator
traces (Jacobi's theta functions).  

This discussion makes also manifest the formal character of the
extra `bulk dimensions' $d_\chi^\perp = 2N+2-d$ of the winding fields
$\chi_\ell$, something already clear from the fact that $d_\chi^\perp$
depends on the number of insertions in the loop. We see that, in those
models with a string-theory embedding, there is a `bulk' codimension
$d_\perp = D-d$,  but its relation with $d_\chi^\perp$ is rather indirect.

{\it iii}) Another deficiency of \fsr, the absence of Koba--Nielsen parameters,
is remedied in \truesr. The interpretation of the full
diagram in the NCFT as a tree-level exchange of $\chi_\ell$-fields was
all right provided we make a further convolution of the vertices with
Feynman parameters. We now understand this feature as a residue of the
full string picture, since Koba--Nielsen parameters map consistently to
Feynman parameters in the SW limit.  Therefore, the $\chi_{\ell}$-field picture
of the NCFT mimics closely the structure of the closed-string channel
in the full string theory.                          

{\it iv}) The sum rule \truesr\ holds for the integrand of the moduli-space
integral. Therefore, it holds independently of the possible occurrence
of open- or closed-channel tachyons in the full string theory. 
This is in contrast with \fsr, which would be invalidated by open-string
tachyons, and perhaps also by closed-string tachyons. It would be very
interesting to study particular examples in detail
 to see the interplay between
the various open/closed tachyons that could appear, including the  
finite temperature Hagedorn tachyon. 
 
{\it v}) Our discussion is tailored to the case of thermal amplitudes.
However, it is clear that the general features  generalize
 to other toroidal compactifications
with various degrees of supersymmetry.

\subsec{An illustrative example}

Unlike Eq. \fsr, the sum rule \truesr\ is valid point by point in the 
$(s,x^{\pm})$ moduli space. As a consequence, it is well defined even for
tachyonic theories for which the integrated expressions would diverge due
to the contribution coming from the moduli space boundaries. This being
so, we can illustrate our sum rule \truesr\ by considering the simplest possible
example and take the two-point function of open-string tachyons on a 
 D$_{d-1}$ 
brane of the $D=26$ critical bosonic string theory.
 In order to avoid unneccesary complications we will consider the static amplitude
where incoming states do not carry time-components of the momenta. 
Thus, the amplitude in the open-string channel can be written as \refs\rkl\refs\rbil
\refs\rlmi
$$
\eqalign{
\CA(p,-p)_{\rm tachyon}=&G_{s}\int_{0}^{\infty}{d\tau\over 2\tau}
(4\pi\alpha'\tau)^{-{d\over 2}}\left[\eta\left({i\tau\over 2\pi}\right)\right]^{-24}
\sum_{\ell\in\bf Z} e^{-{\ell^2\beta^2\over 4\alpha'\tau}}\,\,
e^{-{\alpha'\pi^2\over \tau}p_{\mu}(g^{\mu\nu}-G^{\mu\nu})p_{\nu}}
\cr \times &\, \tau^{2}
\int_{0}^{1}dx^{\pm}\left|2\pi e^{-{\half x_{12}^2}}
{\theta_{2}\left(\left.{ix_{12}\tau\over 2\pi}\right|{i\tau\over 2\pi}\right)
\over \theta_{1}'\left(0\left|{i\tau\over 2\pi}\right.\right)}\right|^{-2}
}
$$
where we have defined $x_{12}\equiv x^{+}-x^{-}$. Comparing this expression with Eqs. 
\ampl\ and \corr\ we can read both $Z(\tau)^{\rm osc}$ and 
$\overline{\CV}(p,x^{\pm},\tau)$, in terms of which the overlap 
\modtrans\ is expressed,
$$
\eqalign{
Z(\tau)^{\rm osc} &= 
\left[\eta\left({i\tau\over 2\pi}\right)\right]^{-24}, \cr
\overline{\CV}(p^{\pm},x^{\pm},\tau) &= e^{{\alpha'\pi^2\over \tau}p_{\mu}G^{\mu\nu}
p_{\nu}}\left|2\pi e^{-{\half x_{12}^2}}
{\theta_{2}\left(\left.{ix_{12}\tau\over 2\pi}\right|{i\tau\over 2\pi}\right)
\over \theta_{1}'\left(0\left|{i\tau\over 2\pi}\right.\right)}\right|^{-2}.
}
$$
Switching from $\tau$ to the closed-string modular parameter $\tau_{2}=2\pi^2/\tau$ and
performing the inversion on the modular functions we find for the partition function 
of the oscillators 
$$
Z\left({2\pi^2\over \tau_2}\right)^{\rm osc} =
\left(\tau_{2}\over \pi\right)^{-12}
\left[\eta\left({i\tau_{2}\over \pi}\right)\right]^{-24},
$$
whereas the function $\overline{\CV}(p^{\pm},x^{\pm},\tau)$ is written
$$
\overline{\CV}\left(p,x^{\pm},{2\pi^2\over \tau_{2}}\right)=
\left({\tau_{2}\over\pi}\right)^{2}
e^{\half \alpha'\tau_{2}p_{\mu}G^{\mu\nu}p_{\nu}}
\left|2\pi{\theta_{4}\left(\left.{x_{12}}\right|
{i\tau_{2}\over \pi}\right)\over 
\theta_{1}'\left(0\left|{i\tau_{2}\over \pi}\right.\right)}
\right|^{-2}.
$$
Using Eq. \modtrans\ we can now obtain the expression for the overlap, namely 
$$
\left\langle {\rm D}_{d-1};V_{p^{+}},x^{+}\left|
e^{-\tau_{2}\Delta_{cl}^{\rm osc}}\right|{\rm D}_{d-1};V_{p^{-}},x^{-}\right\rangle=
e^{\half\tau_{2}}\,\left[\eta\left({i\tau_{2}\over \pi}\right)\right]^{-18}
\left[{\theta_{4}\left({x_{12}}\left|
{i\tau_{2}\over \pi}\right.\right)}\right]^{-2},
$$
where we have used the on-shell condition for the external tachyons, 
$p_{\mu}G^{\mu\nu}p_{\nu}=1/\alpha'$ and also the relation 
$\theta_{1}'(0|\tau)=2\pi\eta^{3}(\tau)$.
Actually, the modular functions can be rewritten
 using their product representations. 
Expanding the resulting
 infinite products in power series of $e^{-2\tau_{2}}$ we 
finally arrive at
\eqn\ex{
\eqalign{
&\left\langle {\rm D}_{d-1};V_{p^{+}},x^{+}\left|
e^{-\tau_{2}\Delta_{cl}^{\rm osc}}\right|{\rm D}_{d-1};V_{p^{-}},x^{-}
\right\rangle
\cr
&= e^{2\tau_{2}}\prod_{k=1}^{\infty}
(1-e^{-2k\tau_{2}})^{-20}\left|1-e^{-(2k-1)\tau_{2}}
e^{2\pi i x^{+}}e^{-2\pi i x^{-}}\right|^{-4} \cr
&= \sum_{n=0}^{\infty}\rho_{\rm D}(n)\,
 C_{n}^{+}(x^{+}) \;e^{-{\half\alpha'} \tau_2 M^2_n}
\;C_{n}^{-}(x^{-})^{*}.
}
}
where $M^2_n = {4\over \alpha'} (n-1)$ is the mass of the level-$n$ oscillator
states, and  $\rho_{\rm D}(n)$ is the level-density of those states with
non-vanishing  coupling to the ${\rm D}_{d-1}$
brane.
Notice that, because of the structure of the
product representation for the modular functions, the coefficient of $e^{-2(n-1)\tau_{2}}$
in the series always factorizes into  contributions from the two different boundaries, 
$C_n^{\pm}(x^{\pm})$, weighted by the level-density, $\rho_{\rm D}(n)$. 
Comparing this expression with the sum rule 
\truesr\ we read-off
 the couplings of a level-$n$ closed-string state $|\Psi_{n}\rangle$ 
to the boundary state with an external open-string tachyon insertion
\eqn\cous{
\left\langle {\rm D}_{d-1};V_{p^{+}},x^{\pm}{\Big |}\Psi_{n}\right\rangle
= C_{n}^{\pm}(x^{\pm}).
}
In this example we explicitely see how the coupling between $|\Psi_{n}\rangle$ and
the boundary state with a tachyon insertion is in general nonvanishing 
for all values of $n$ and thus {\it all} closed-string oscillator
levels run in the cylinder. 
As a consequence, the effective $\chi_{\ell}$-fields cannot be seen as some undecoupled
winding string state in the SW limit, but rather as
 a superposition of all massive 
closed-string modes with coherent couplings to the elementary quanta of the NCFT.

\newsec{Concluding remarks}

In the present paper we have tried to identify the stringy connection of the
recently conjectured winding modes emerging in NCFT \refs\rtexas.
We have seen how a thermal loop in NCFT 
can be represented in a `dual channel' picture as an infinite tower of 
tree-level exchanges of some effective fields $\chi_{\ell}$ with 
masses proportional to $|\beta\ell|$ ($\ell\in {\bf Z}$) and kinetic term
$\partial \circ \partial$ in the effective action. The scaling of the masses
of these fields with the length of the euclidean time suggests a winding mode
interpretation  for them. 

In many respects these fields are similar to the $\psi$-fields introduced in
Refs. \refs\rp\refs\rrase. It is important to notice however that there are a number of
differences. First of all, in the thermal case we have not just one, but 
an infinite tower of effective fields replacing the thermal loop in nonplanar
amplitudes. As a consequence, one is able to replace the {\it whole} thermal loop
by a tree-level exchange of these fields and not just the high energy part as in 
\refs\rp. Most importantly, the $\chi_{\ell}$-fields have nonstandard Feynman rules.
If the nonlocality in NCFT is just encoded in a nonpolynomial dependence of the interaction 
vertices on the
 incoming momenta, in the perturbation theory for the effective
 $\chi_{\ell}$-fields 
 the interaction vertices are convoluted in finite dimensional integrals 
over the Feynman parameters of the original diagram.

Actually, both features, winding-like masses and integration over the 
relative moduli of the interaction vertices, strongly suggest a stringy interpretation.
We have found that such interpretation exists in those models which can be obtained from
a D-brane theory in the presence of a constant $B$-field in the SW limit\foot{The
phenomenon of UV/IR mixing is also present in 
nonrelativistic noncommutative field theories for which no obvious embedding into
a string theory seems to exist \refs\rgll.}.
In this case, we found that the scaling of the masses of the $\chi_{\ell}$-fields
with the length of the thermal circle is a residue of winding closed-string states, whereas
the convolution over the vertex moduli descend from the integration over the Koba--Nielsen
moduli of the D-brane boundary states. 

It is however important to notice that the `winding states' identified in \refs\rtexas\
have no simple interpretation in terms of {\it individual} string states that fail to decouple 
in the SW limit. On the contrary, the $\chi_{\ell}$-fields have to be considered formal
devices to represent a coherent coupling of an infinite number of closed-string states.
These fields have effective couplings to the ordinary fields that can be derived from the
elementary coupling of closed strings to D-brane boundaries via `sum rules' 
involving the full tower of closed-string oscillator modes in the bulk. 
This is a rather unusual picture, and essentially is telling us that the SW limit 
is not an ordinary low-energy limit in the closed-string channel since {\it all}
massive states are squeezed below the gap of the winding modes.
As a consequence, the resulting NCFT present a degenerate version of the open/closed
string duality of the original string theory: the ordinary `open' representation of
the Feynman diagram in NCFT and the `closed' dual channel in terms of the winding 
$\chi_{\ell}$-fields.

\vskip 0.3cm

\noindent {\bf Acknowledgements}
\vskip0.2cm 
\noindent
It is a pleasure to thank Luis Alvarez-Gaum\'e, Jan de Boer,
Karl Landsteiner, Esperanza L\'opez,  Niels Obers and Eliezer Rabinovici
for useful discussions. G.A. warmly thanks the Spinoza Instituut and specially 
Gerard 't Hooft for hospitality. J.G. is partially supported by CERN and by 
Spanish Science Ministry
Grant AEN98-043 and Catalan Science Foundation Grant GC 1998 SGR.  
The work of M.A.V.-M. has been supported by FOM ({\it Fundamenteel Onderzoek van de Materie}) 
and University of the Basque Country Grants UPV 063.310-EB187/98 and UPV 172.310-G02/99 and
Spanish Science Ministry Grant AEN99-0315. M.A.V.-M. wishes also 
to thank CERN Theory Division for hospitality during the completion of this work.

\listrefs

\bye